# Reduced-Alphabet QUBO/Ising Formulation for Constraint-Driven Cyclic Peptide Sequence Design

Yan Zhou[1,*], Yuqi Wang[1], Xin Li[1]

[1]School of Life Sciences, Fudan University, Shanghai 200438, China

Correspondence: Yan Zhou (zhouy@fudan.edu.cn)

# Abstract

Cyclic peptide design requires balancing local residue preferences with constraints from ring-forming chemistry, residue spacing, topology, target compatibility, and developability. Here, we present a reduced-alphabet quadratic unconstrained binary optimization (QUBO)/Ising formulation for constraint-driven cyclic peptide sequence design. Amino acids are grouped into physicochemical or interaction-based residue classes, and peptide positions are represented by binary residue-class assignment variables. The objective combines one-hot sequence validity, cyclization constraints, optional target-compatibility terms, motif and composition rules, and coarse developability proxies. By modifying the relevant constraint terms, the same framework can represent head-to-tail, disulfide-bridged, stapled, and bicyclic peptide designs. A resource-aware eight-class alphabet motivated by MJ interaction-profile clustering is used as a default representation to balance coarse interaction-pattern preservation with encoding cost. The resulting QUBO/Ising objective is solver-agnostic and can be explored using classical or quantum-compatible binary optimization procedures. The model is intended as an early-stage search-space reduction and prioritization layer: it produces low-energy residue-class sequences rather than final molecular candidates, which require amino-acid decoding, cyclization-aware construction, and downstream structural or experimental validation.

# 1. Introduction

Peptide therapeutics are well suited to drug-discovery problems involving biological interfaces that are too large, shallow, or dynamic for many conventional small molecules.

Their modular sequences and chemical accessibility make them attractive scaffolds for residue substitution, incorporation of non-natural amino acids, backbone or side-chain modification, and peptidomimetic redesign, thereby enabling systematic tuning of binding affinity, target selectivity, and pharmacological properties [1]. Liraglutide, a modified glucagon-like peptide-1 receptor agonist used for type 2 diabetes, illustrates how peptide analogues can be optimized to improve therapeutic performance and clinical utility [2]. Enfuvirtide, a synthetic 36-amino-acid HIV-1 fusion inhibitor, blocks viral entry by targeting the gp41-mediated membrane-fusion process, showing that peptide drugs can engage protein-mediated biological interfaces that are difficult to modulate with many conventional small molecules [3]. Despite these advantages, many linear peptides remain limited by poor metabolic stability, susceptibility to proteolytic degradation, short circulation time, and unfavorable pharmacokinetic behavior [1]. These limitations have motivated structural strategies that preserve peptide-like molecular recognition while improving stability and drug-like performance.

Cyclization is one of the most direct strategies for addressing the pharmacokinetic liabilities of linear peptides. By connecting distant positions within a peptide chain, it can reduce terminal exposure, restrict backbone motion, decrease conformational heterogeneity, and increase resistance to enzymatic degradation [4]. The resulting structural preorganization may also favor bioactive conformations and improve developability-related properties, including metabolic stability and, in selected cases, oral absorption [4]. Clinically successful cyclic or conformationally constrained peptide drugs illustrate these advantages. Cyclosporine A, a cyclic peptide immunosuppressant, provides a classic example in which cyclic topology and other chemical features contribute to proteolytic stability, oral availability, and clinical utility [4,5]. Octreotide, a cyclic somatostatin analogue, was developed to overcome the very short in vivo half-life of native somatostatin and remains a clinically useful example of stability-enhanced peptide analogue design [4,6]. In this sense, the advantages of cyclic peptides arise from both amino acid composition and ring-imposed geometry. Such constraints have been associated with improved binding affinity, enhanced in vivo stability, and improved drug-like behavior compared with many unconstrained linear peptide formats [4].

The same ring constraints that improve peptide properties also make cyclic peptide design substantially more constrained. Ring formation changes molecular topology and introduces requirements that must be satisfied at both sequence and geometric levels [7]. Different cyclization modes impose different design rules. N-to-C terminal cyclization requires compatible terminal chemistry, appropriate ring size, and favorable pre-cyclization conformations; side-chain-to-side-chain cyclization can stabilize secondary structures such as α-helices and β-sheets, giving rise to stapled peptides [7]. Disulfide cyclization requires cysteine residues at suitable positions, as illustrated by oxytocin, a nonapeptide containing a disulfide bridge between two cysteines [8]. Stapled peptides impose additional side-chain identity and spacing requirements; ALRN-6924, for example, is a stabilized α-helical stapled peptide that mimics the N-terminal domain of p53 and binds MDM2 and MDMX [9]. A candidate cyclic peptide therefore cannot be judged only by local residue preferences or statistics learned from linear chains. It must also satisfy global closure, residue-type, spacing, and spatial-compatibility constraints. The feasible design space is consequently smaller and more structured than that of unconstrained linear peptide generation. This helps explain why linear peptide generation methods do not transfer directly to cyclic peptides, and why post-filtering or hard-coded strategies can suffer from low acceptance rates or limited generality across cyclization patterns [10].

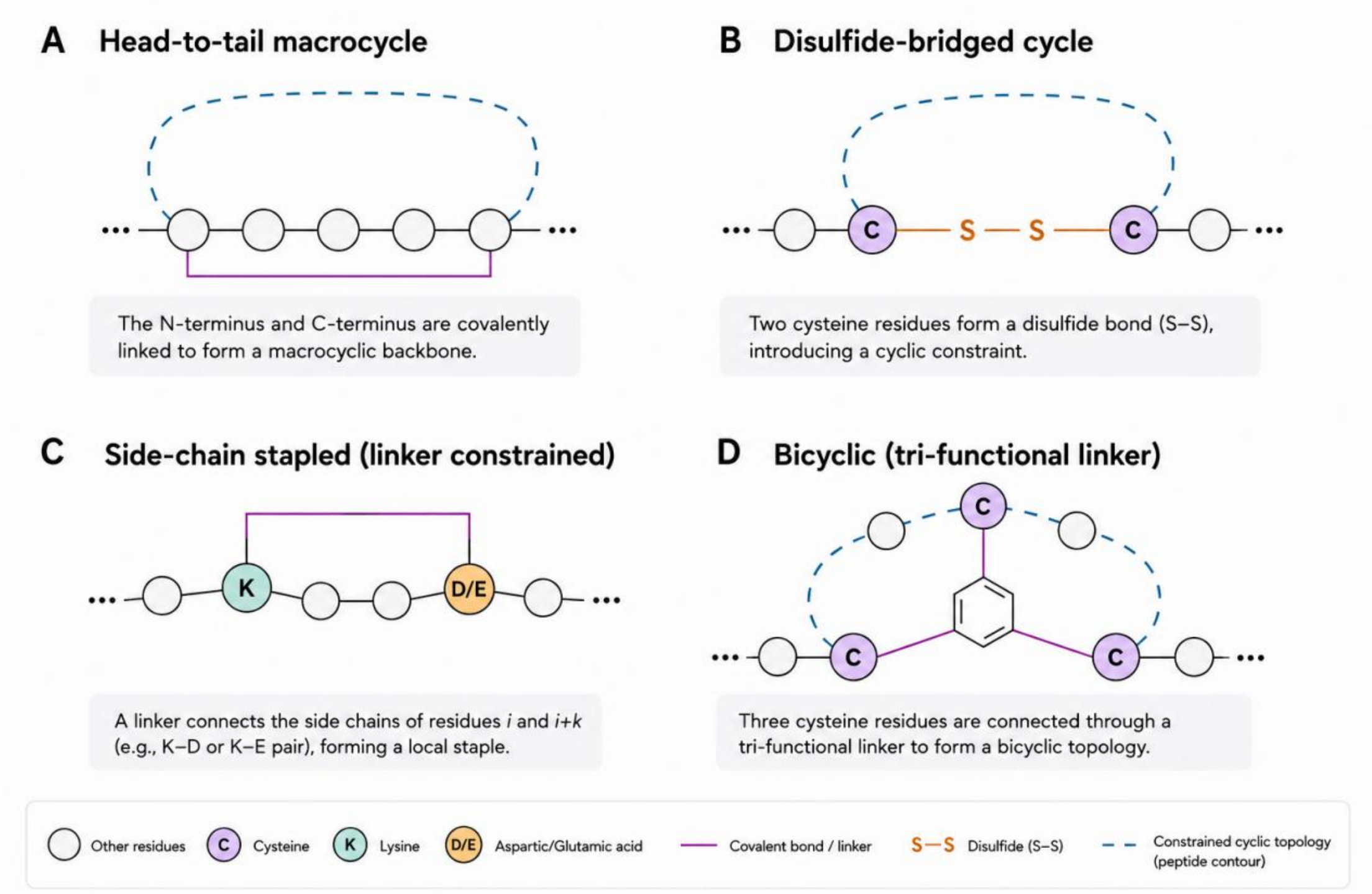

**Figure 1. Representative cyclization modes in cyclic peptide design.** (A) Head-to-tail macrocyclization links the N-terminus and C-terminus to form a macrocyclic backbone. (B) Disulfide-bridged cyclization forms an S – S bond between two cysteine residues. (C) Side-chain stapling introduces a local covalent constraint through a linker connecting selected side chains. (D) Bicyclic cyclization connects three cysteine residues through a trifunctional linker to form a bicyclic topology. These cyclization modes impose distinct residue-type, positional, and topological constraints, motivating a unified constraint-based formulation for cyclic peptide design.

Recent AI-based peptide design has shifted part of the field from scoring pre-enumerated candidates toward conditional generation. For linear peptides, target-conditioned language-model approaches such as PepMLM can generate de novo peptide binders from a protein target sequence by masking and reconstructing the binder region, allowing target-specific peptide ligands to be proposed without structural input [11]. For cyclic peptides, CP-Composer represents a complementary direction: it formulates cyclic peptide design as a zero-shot constrained generation task and decomposes cyclization patterns into elementary type and distance constraints that can be combined to represent stapled, head-to-tail, disulfide, and bicyclic strategies [10]. For the present problem, the main implication is that peptide generation can be guided by explicit design conditions rather than left entirely to post-generation filtering. Cyclic peptide generation therefore requires topology and geometry constraints that are absent from unconstrained linear peptide design.

Quantum computing offers a related but distinct line of work in peptide and protein modeling, and its applications can be broadly divided into two routes. The first route is conformation-oriented optimization. In this setting, a fixed peptide or protein sequence is encoded into lattice walks, contact variables, torsional states, or other discrete representations, with the objective of sampling low-energy conformations or docking poses. This direction can be viewed as a long-term technical route toward increasingly larger folding and structure-prediction problems, although near-term implementations remain limited by qubit resources, circuit depth, noise, and simplified energy models. QAOA has been investigated for low-energy conformational sampling of short peptides, but deep circuits were required for accurate low-energy sampling and random sampling could match QAOA performance up to a small overhead in simplified settings [12]. The second route is design-oriented or generative optimization, where the aim is to generate peptide candidates under target-, topology-, or property-related constraints. AI models such as PepMLM and CP-Composer exemplify this conditional-generation view for linear and cyclic peptides, respectively [10,11]. A quantum-compatible counterpart is quantum-assisted de novo peptide binder design, in which reduced amino-acid encodings, sequence variables, and coarse-grained binding poses are optimized within a multiscale classical–quantum framework [13]. Cyclic peptides have also appeared in QUBO-based work, but in a different setting: lattice-based cyclic peptide docking has been formulated as a QUBO and constraint-programming problem, where the input is a given cyclic peptide and the objective is to place it near a protein active site under both cyclization and peptide–protein interaction constraints [14]. These studies support the relevance of quantum-compatible optimization for peptide problems, but their scope remains fixed-sequence conformational sampling, linear or protein-binding peptide design, or docking of a given cyclic peptide. They leave unresolved how cyclic peptide sequence- or class-level generation can be formulated directly as a QUBO/Ising optimization problem.

A useful quantum-compatible peptide model is therefore not defined simply by choosing a quantum device or solver. The first pressure point is encoding. In conformation-oriented models, the encoded variables usually describe lattice walks, turn states, contacts, torsional states, or other discrete representations of molecular geometry; in design-oriented models, the encoded variables describe amino-acid identities, residue classes, or sequence-position assignments. Both routes face the resource constraints of the noisy intermediate-scale quantum era, where limited qubit number, device connectivity, noise, circuit depth, and embedding overhead restrict the size of peptide problems that can be directly encoded. This is why reduced-alphabet and coarse-grained representations are attractive for near-term peptide optimization. In sequence-design settings, directly encoding all 20 natural amino acid identities at every peptide position can rapidly increase the number of binary variables and interaction terms. Reduced-alphabet representations address this issue by grouping amino acids into fewer physicochemical families, thereby lowering the size of the resulting optimization problem; such reduced molecular encodings have been used in quantum-assisted peptide binder design [13]. Encoding alone, however, does not define the model. The objective function determines how sequence preferences, target-binding scores, cyclization constraints, motif requirements, compatibility rules, and drug-like property terms are converted into quadratic penalties or rewards [10,13,14]. Prior cyclic peptide docking work is particularly relevant in this respect because it encodes ring closure and peptide–protein interaction as distinct optimization components, suggesting that topology-related and target-related requirements can be represented as separate terms in the objective function [14]. Scoring introduces another layer of approximation. MJ-type residue–residue contact potentials provide a convenient coarse-grained source of interaction energies, but such knowledge-based statistical potentials should be treated as one possible scoring component rather than as a complete cyclic-peptide force field [15].

After the Hamiltonian is specified, the remaining choices concern how the model is solved and how its output is read. The same QUBO/Ising objective can be processed through classical annealing, quantum annealing, variational circuits, QAOA-like sampling, constraint programming, or other Hamiltonian-based optimization procedures; the circuit implementation is therefore not intrinsic to the peptide model but depends on the selected algorithmic route [12–14]. A design-oriented D-Wave example compared a quantum annealer with a classical optimizer for de novo protein-binding peptide design [13]. By contrast, cyclic peptide docking QUBO instances have been solved with classical simulated annealing, highlighting both the relevance and the scaling limitations of quantum-amenable formulations for cyclic peptide problems [14]. The readout mode also matters. Estimator-like workflows evaluate observable expectation values, whereas sampler-like workflows return measurement bitstrings that can be post-processed into candidate molecular configurations or sequences [16]. For reduced-alphabet cyclic peptide class generation, the sampler-like view is especially natural: low-energy bitstrings can be decoded directly into residue-class assignments.

What remains missing is a formulation that treats cyclic peptide generation itself as a reduced binary optimization problem. Recent AI-based cyclic peptide generation has shown that cyclization can be expressed through composable type and distance constraints, but it remains a neural generative framework based on learned geometric representations and available peptide/protein–peptide structural data [10]. CP-Composer addresses the scarcity of cyclic peptide data by learning unit constraints from available linear peptide data and composing them at inference time, but it is not designed as a quantum-compatible binary optimization model [10]. Quantum-assisted peptide design has demonstrated the value of reduced molecular encodings and binary optimization, yet existing work has focused on protein-binding peptide design with joint sequence and binding-pose optimization rather than cyclic peptide class-sequence generation under explicit cyclization constraints [13]. Prior QUBO/CP work on cyclic peptides is also informative because it incorporates both cyclization and peptide–protein interaction constraints into a docking formulation, but its task is to dock a given cyclic peptide rather than to generate cyclic peptide sequences or residue-class patterns from design constraints [14]. We therefore propose a reduced-alphabet QUBO/Ising formulation for cyclic peptide residue-class sequence design, in which peptide positions are represented by reduced residue-class assignments and design requirements are encoded as separate objective terms. These objective terms cover sequence validity, class preference, cyclization feasibility, target-binding compatibility, motif and composition requirements, sequence compatibility, and property-related constraints. Rather than claiming quantum advantage or replacing downstream structural validation, this work establishes a quantum-compatible theoretical formulation that connects cyclic peptide design rules with combinatorial optimization.

# 2. Problem Definition and Modeling Scope

This work considers cyclic peptide design at the reduced residue-class level. The design object is a cyclic peptide of length N, but the model does not directly optimize a full atomistic peptide structure or a complete 20-amino-acid sequence. Instead, each peptide position is assigned to one residue class from a reduced alphabet $C$. A residue class may represent a physicochemical group, such as hydrophobic, polar, charged, cysteine-like, proline/glycine-like, or linker-compatible residues. The resulting design problem is therefore a class-sequence generation problem rather than a direct structure-prediction or docking problem.

The motivation for this abstraction is twofold. First, cyclic peptide design is more constrained than unconstrained linear peptide generation because ring formation imposes residue-type, spacing, topology, and compatibility requirements. Recent cyclic peptide generation work has shown that different cyclization modes can be expressed through composable type and distance constraints, including stapled, head-to-tail, disulfide, and bicyclic strategies [10]. Second, quantum-compatible peptide design is strongly affected by the size of the encoded search space. Directly representing all natural amino acids at every position increases the number of binary variables and pairwise interaction terms. Reduced molecular encodings have therefore been used in quantum-assisted peptide binder design to make peptide sequence optimization more compact [13].

The inputs to the proposed formulation are a peptide length $N$, a reduced residue-class alphabet $C$, one or more cyclization rules, optional target-binding priors, and additional user-defined design rules. Cyclization rules specify which positions or position pairs are relevant for a selected ring-formation strategy. For example, disulfide cyclization may require cysteine-like classes at two positions, head-to-tail cyclization may involve compatibility between terminal positions, and stapled peptides may impose linker-compatible classes at positions separated by a design-dependent interval. Target-binding priors describe the binding object as a fixed external environment. These priors may include target-facing peptide positions, binding-site residue classes, coarse contact weights, interface annotations, or structural hypotheses. Prior QUBO work on cyclic peptide docking is relevant here because it treats cyclization and peptide–protein interaction constraints as distinct components of a single optimization model [14]. In the present formulation, however, the task is not docking a given cyclic peptide, but generating residue-class patterns that are compatible with both cyclization and target-related design requirements.

The optimized variables belong only to the peptide. For each peptide position $i$ and residue class $c$, the model introduces a binary class-assignment variable $x_{i,c}$. A valid class sequence requires each peptide position to select exactly one class. After optimization, a low-energy bitstring can be decoded into a residue-class sequence

$$z = (z_1,\ldots,z_N), \qquad z_i \in C.$$

This decoded sequence is the primary output of the model. It is not yet a final chemical candidate. Concrete amino-acid sequences, atomistic cyclic conformers, docking poses, molecular-dynamics trajectories, and experimental candidates are treated as downstream outputs. This distinction is important because the present work formulates an early-stage design-space reduction model rather than a complete cyclic peptide discovery pipeline.

The reduced alphabet can be chosen in different ways. A simple implementation may group amino acids by physicochemical properties. A more data-driven implementation may cluster amino acids according to MJ-type residue–residue interaction profiles, using knowledge-based contact potentials as a coarse source of interaction information [15]. In this view, the reduced alphabet is not intended to erase amino-acid specificity completely, but to compress the design space while preserving interaction patterns that are useful at the class-sequence level. An eight-class alphabet is especially attractive as a default setting because it can be represented by three compact binary variables per position in a hardware-aware implementation. However, the main formulation below uses one-hot class indicators because they keep class preferences, pairwise compatibility terms, and constraint penalties explicitly quadratic. Compact encodings are therefore treated as implementation-level variants that may require quadratization or low-order approximation when arbitrary class-score lookup tables are used.

The mathematical form of the model follows a standard penalty-Hamiltonian view of constrained binary optimization. In such formulations, hard validity conditions and softer design preferences are written as energy terms, and candidate solutions are ranked by their total energy. Ising and QUBO formulations have been widely used as general languages for expressing discrete optimization problems, where constraints can be enforced by penalty terms and the resulting binary objective can be mapped to spin variables when needed [17]. This perspective is useful for cyclic peptide design because sequence validity, cyclization feasibility, target compatibility, motif requirements, composition preferences, and coarse developability filters can all be represented as modular contributions to one objective function.

Several exclusions define the scope of the present model. The formulation does not directly predict three-dimensional cyclic peptide structures, does not perform peptide–protein docking, does not estimate binding free energy, and does not simulate the chemical mechanism of ring closure. It also does not claim quantum advantage. The QUBO/Ising form makes the model compatible with classical heuristics, simulated annealing, quantum annealing, QAOA-like samplers, or other binary optimization procedures, but the choice of solver is separate from the peptide model itself. The purpose of the present section is therefore to define the design problem and its boundaries; the next section gives the explicit QUBO/Ising formulation used to encode this reduced-alphabet cyclic peptide design task.

# 3. Reduced-Alphabet QUBO/Ising Formulation

## 3.1 Residue-class variables and sequence validity

We consider a cyclic peptide of length N and a reduced residue-class alphabet $C = \{1,\ldots,K\}$.

For clarity, the main symbols used in this formulation are summarized below.

**Table 1. Main notation used in the reduced-alphabet QUBO/Ising formulation.**

| Symbol | Meaning |
| --- | --- |
| $N$ | length of the cyclic peptide |
| $C = \{1,\ldots,K\}$ | reduced residue-class alphabet |
| $K$ | number of residue classes |
| $x_{i,c}$ | binary indicator that position (i) is assigned to class (c) |
| $z = (z_1,\ldots,z_N)$ | decoded residue-class sequence |
| $T_{cyc}$ | set of cyclization-related type constraints |
| $\mathcal{E}_{cyc}$ | set of cyclization-related position pairs |
| $\mathcal{B}$ | target-facing peptide positions |
| $S$ | target binding-site residues or sites |
| $y_{r,d}$ | fixed class indicator for target site (r) |
| $\lambda$ | penalty weight |

| Symbol | Meaning |
|---|---|
| $C_{ir}$ | fixed contact weight between peptide position i and target site r |
| $C_{target}$ | reduced residue-class alphabet for target residues or sites |

Each class may represent a physicochemical group, such as hydrophobic, polar, positively charged, negatively charged, cysteine-like, proline/glycine-like, or linker-compatible residues. The specific alphabet is not fixed by the formulation; it can be chosen according to the design task, the desired chemical resolution, MJ-type interaction-profile clustering, and the available downstream decoding strategy [15].

For each peptide position and residue class, we introduce a binary variable

$$x_{i,c} \in \{0,1\}, \qquad i = 1,\ldots,N, \quad c \in C.$$

Here, $x_{i,c} = 1$ indicates that position (i) is assigned to residue class (c). A valid residue-class sequence requires exactly one class to be selected at each position. This is enforced by a one-hot penalty,

$$H_{onehot} = \sum_{i=1}^{N} \left(1 - \sum_{c \in C} x_{i,c}\right)^2.$$

This form is a standard assignment-style penalty in QUBO/Ising formulations [17]: a multi-class choice is represented by binary indicators, and invalid assignments are penalized energetically. The penalty is zero only when every peptide position is assigned to exactly one residue class. If no class or multiple classes are selected at a position, the term contributes a positive energy penalty. This block-wise one-hot encoding follows the same general logic as reduced molecular encodings used in quantum-compatible peptide design, but here the variables represent cyclic peptide residue classes rather than full amino-acid identities or binding-pose states [13].

A candidate solution of the QUBO model is therefore a bitstring over all variables $x_{i,c} \mid i = 1,\ldots,N$ and $c \in C$. Once the one-hot constraint is satisfied, the bitstring can be decoded into a residue-class sequence

$$z = (z_1,\ldots,z_N), \; z_i = \arg\max_{c \in C} x_{i,c}.$$

where $z_i$ is the unique class selected at position i. Equivalently, under the one-hot constraint, $z_i$ denotes the residue class for which $x_{i,c} = 1$.

This class-level sequence is the primary output of the formulation. Concrete amino-acid sequences, three-dimensional conformers, docking poses, and experimental candidates are treated as downstream outputs rather than as variables in the present model. The one-hot class-indicator representation is used here as a logical QUBO formulation because it keeps class preferences, pairwise compatibility terms, and cyclization constraints explicitly quadratic. In hardware-aware implementations, a compact encoding may also be used; for example, an eight-class alphabet can be represented by three binary variables per peptide position. Such compact encodings reduce the number of logical variables, but arbitrary class-score lookup tables may introduce higher-order Boolean terms that require quadratization or low-order approximation. We therefore use the one-hot form for the main formulation and treat compact encodings as an implementation-level variant.

## 3.2 Composable cyclization constraints

The one-hot term defines a valid residue-class sequence, but it does not by itself distinguish a linear peptide from a cyclic peptide. Cyclic peptide design requires additional constraints because ring formation imposes residue-type, positional, and pairwise compatibility requirements. Different cyclization strategies, such as disulfide bridging, head-to-tail cyclization, side-chain stapling, or bicyclic linkage, impose different chemical rules, but these rules can be represented in a common constraint-based form.

Inspired by composable constraint formulations in cyclic peptide generation, we decompose cyclization requirements into two basic classes of constraints [10]: type-like constraints and distance-inspired pairwise constraints. Type-like constraints specify that selected peptide positions must belong to residue classes compatible with a given cyclization chemistry. Distance-inspired constraints specify that two positions must form a cyclization-compatible pair at the class-sequence level. In the present formulation, these constraints are not implemented as geometric diffusion conditions or explicit Euclidean distances. They are instead reduced-alphabet compatibility proxies: they specify whether two residue-class assignments are plausible for the intended cyclization chemistry.

Let $T_{cyc}$ denote the set of cyclization-related type constraints. Each element $((i,c^*) \in T_{cyc})$ indicates that position i is required to take residue class $c_i^*$. The corresponding penalty is

$$H_{type} = \sum_{(i,c^*) \in T_{cyc}} \left(1 - x_{i,c^*}\right)^2 .$$

This term is zero when all specified positions satisfy their required residue-class assignments. It becomes positive when a required class is absent. For example, a disulfide constraint may require two positions to be assigned to a cysteine-like class, whereas a stapled peptide may require two positions to be assigned to linker-compatible classes. A bicyclic peptide can be represented by adding multiple type constraints, such as several cysteine-like or linker-reactive positions.

Type constraints alone do not capture whether two selected positions form a chemically or topologically compatible pair. We therefore define a set of cyclization-related position pairs,

$$\mathcal{E}_{\mathrm{cyc}} \subseteq \{(i,j):1 \le i < j \le N\}.$$

For each pair $(i,j) \in \mathcal{E}_{\mathrm{cyc}}$, we assign a pairwise compatibility score $\Gamma_{ij}^{cd}$ for selecting residue class $c$ at position $i$ and residue class $d$ at position $j$. This gives the pairwise cyclization penalty

$$H_{\mathrm{pair}} = \sum_{(i,j)\in \mathcal{E}_{\mathrm{cyc}}} \sum_{c,d\in C} \Gamma_{ij}^{cd} x_{i,c} x_{j,d}.$$

Here $\Gamma_{ij}^{cd}$ can be chosen as a penalty or reward depending on whether the class pair $(c,d)$ supports the desired cyclization pattern. For incompatible class pairs, $\Gamma_{ij}^{cd}$ is assigned a positive penalty. For compatible class pairs, the value may be set to zero or to a favorable negative score. Because the present model does not include atomic coordinates, this term should not be interpreted as an explicit geometric distance constraint. It is a reduced-alphabet compatibility proxy that encodes whether two residue-class assignments are plausible for the intended cyclization chemistry.

Several common cyclization modes can be written using this notation. For a disulfide-bridged peptide with two specified positions $p$ and $q$, the required cysteine-like assignments can be written as

$$H_{SS} = (1 - x_{p,c_{Cys}})^2 + (1 - x_{q,c_{Cys}})^2.$$

This form specifies the required residue classes for the disulfide-forming positions. If the positions are not fixed in advance, a related global count constraint can be used later in the motif or composition term to require a specified number of cysteine-like positions.

For head-to-tail cyclization, the relevant pair is the first and last peptide positions. A simple class-level compatibility term is

$$H_{HT} = \sum_{c,d \in C} \Gamma_{1N}^{cd} x_{1,c} x_{N,d}.$$

Here $\Gamma_{1N}^{cd}$ encodes whether the residue classes assigned to the N- and C-terminal positions are compatible with the desired head-to-tail closure. This term does not model the full atomistic amide-bond formation process; it only biases the class sequence toward terminal assignments compatible with downstream cyclization.

For stapled peptides, a cyclization rule can specify two positions separated by a design-dependent interval $k$, such as $(i, i+3)$, $(i, i+4)$ or $(i, i+7)$, depending on the intended secondary-structure geometry and linker chemistry. A staple constraint can combine type requirements with pairwise compatibility:

$$H_{staple} = \left(1 - x_{p,c_{link}^{(1)}}\right)^2 + \left(1 - x_{p+k,c_{link}^{(2)}}\right)^2 + \sum\nolimits_{c,d \in C} \Gamma_{p,p+k}^{cd} x_{p,c} x_{p+k,d}$$

where $c_{link}^{(1)}$ and $c_{link}^{(2)}$ denote linker-compatible residue classes. More detailed staple rules can be represented by assigning different linker-compatible classes to the two positions or by modifying the pairwise compatibility coefficients.

The total cyclization Hamiltonian is then written as a modular sum,

$$H_{cyclization} = \omega_{type} H_{type} + \omega_{pair} H_{pair} + \omega_{topo} H_{topology}.$$

Here $H_{topology}$ denotes optional topology-level penalties that enforce design-specific rules beyond individual type and pairwise terms, such as the number of cyclization links, allowed linker patterns, or incompatibility between multiple simultaneous cyclization modes. The weights $\omega_{type}$, $\omega_{pair}$ and $\omega_{topo}$ control the relative strength of each constraint class within $H_{cyclization}$.

This construction allows different cyclization strategies to be represented by adding, removing, or reweighting constraint terms rather than redefining the entire model. Disulfide, head-to-tail, stapled, bicyclic, and multi-cyclized peptides can therefore be treated within the same QUBO framework. The formulation remains intentionally coarse-grained: it constrains cyclization feasibility at the residue-class level, while atomistic ring closure, conformer generation, and structural validation are deferred to the downstream design workflow.

## 3.3 Target-binding compatibility constraints

Cyclization constraints restrict the candidate sequence to patterns compatible with ring formation, but they do not specify whether the resulting cyclic peptide is suitable for a particular binding target. For target-aware cyclic peptide design, the QUBO objective should therefore include an additional term that evaluates how residue-class assignments at selected peptide positions match the chemical environment of a target binding region [13,14]. This term provides a coarse residue-class compatibility score for target-facing peptide positions, using target priors to bias sequence generation before structure-based validation.

Let $\mathcal{B} \subseteq \{1,\ldots,N\}$ denote the set of peptide positions assumed or predicted to face the target. These positions may be specified from prior structural knowledge, motif information, interface annotations, docking hypotheses, or user-defined design assumptions. For each binding-facing position $i \in \mathcal{B}$ and residue class $c \in C$, we define a target compatibility coefficient $T_{i,c}$. A simple linear target-binding compatibility term can then be written as

$$H^{(1)}_{\mathrm{target}} = \sum_{i \in \mathcal{B}} \sum_{c \in C} T_{i,c}\, x_{i,c}.$$

Here $T_{i,c}$ can be interpreted as a penalty or reward for assigning residue class $c$ to target-facing peptide position $i$. A lower value favors the assignment, whereas a higher value penalizes it. For example, if a region of the target is negatively charged, positively charged peptide residue classes may receive favorable scores at positions expected to contact that region. If a target subpocket is hydrophobic, hydrophobic or aromatic classes may be favored. Conversely, residue classes that are sterically or chemically incompatible with a local target environment can be assigned positive penalties.

A more explicit form can be obtained by representing the target binding region in the same reduced-class language. Let S denote a set of target residues or target surface sites relevant to peptide binding. To avoid assuming that every target-facing peptide position interacts equally with every target site, we introduce a fixed contact weight ($C_{ir} \geq 0$) between peptide position ($i \in \mathcal{B}$) and target site ($r \in S$). This weight can be binary or continuous and may come from a hypothesized contact map, binding-site annotation, coarse docking hypothesis, or user-defined design prior. For each target site $r \in S$, let

$$y_{r,d} \in \{0,1\}, \qquad d \in C_{\mathrm{target}},$$

indicate whether target site r belongs to target residue class $d$. Unlike $x_{i,c}$, the variables $y_{r,d}$ are fixed inputs derived from the known target sequence, structure, or annotated binding site; they are not optimized. Pairwise peptide-target compatibility can then be written as

$$H^{(2)}_{\mathrm{target}} = \sum_{i \in \mathcal{B}} \sum_{r \in S} C_{ir} \sum_{c \in C} \sum_{d \in C_{\mathrm{target}}} W^{\mathrm{target}}_{cd}\, x_{i,c} y_{r,d}.$$

The coefficient $W^{\mathrm{target}}_{cd}$ represents class-pair compatibility, whereas $C_{ir}$ encodes the presumed contact strength or relevance between peptide position i and target site r. Because $y_{r,d}$ is fixed, this term remains a QUBO-compatible contribution over the peptide variables $x_{i,c}$. If target-site classes are known, the expression can be reduced to an effective linear score,

$$T_{i,c}^{eff} = \sum_{r \in S} C_{ir} \sum_{d \in C_{target}} W_{cd}^{target} y_{r,d},$$

so that

$$H_{target}^{(2)} = \sum_{i \in \mathcal{B}} \sum_{c \in C} T_{i,c}^{eff} x_{i,c}.$$

This form is useful when the target is treated as a fixed external environment rather than as part of the optimization problem. It allows target information to bias peptide class generation without introducing additional optimized variables for the target.

The target-binding term can also include pairwise dependencies between peptide positions when the desired binding pattern depends on the joint arrangement of multiple residue classes. For example, two peptide positions may need to form a complementary charged pair, preserve a hydrophobic face, or avoid an incompatible class combination at neighboring target-contacting positions. Such effects can be represented by

$$H_{target}^{(3)} = \sum_{\substack{i<j \\ i,j \in \mathcal{B}}} \sum_{c,d \in C} R_{ij}^{cd} x_{i,c} x_{j,d},$$

where $R_{ij}^{cd}$ penalizes or rewards a pair of residue-class assignments among target-facing peptide positions. This term remains quadratic in the peptide variables and therefore fits naturally into the QUBO formulation.

Combining these options, the target-binding compatibility contribution can be written as

$$H_{target} = \omega_{lin} H_{target}^{(1)} + \omega_{pt} H_{target}^{(2)} + \omega_{tpair} H_{target}^{(3)}.$$

The weights $\omega_{lin}$, $\omega_{pt}$, and $\omega_{tpair}$ determine how strongly each level of target information contributes within $H_{target}$. In a minimal implementation, only the linear term $H_{target}^{(1)}$ may be used. If target residue classes or an approximate contact map are available, the peptide-target term $H_{target}^{(2)}$ can be included. If class-level patterns across multiple peptide positions are important, $H^{(3)}_{target}$ can be added.

This target term is deliberately less detailed than docking. It does not predict an atomistic binding pose, does not evaluate conformational entropy, and does not replace downstream structural validation. Its role is to incorporate target preference into the same binary optimization problem that also contains sequence-validity and cyclization constraints. The model therefore generates cyclic peptide class sequences that are not only compatible with the selected ring-formation strategy, but also biased toward a specified target-binding environment.

## 3.4 Motif, composition, and developability terms

In addition to cyclization and target-binding compatibility, cyclic peptide design often requires lower-level sequence rules that reflect known motifs, residue composition, or coarse developability preferences. These terms act as coarse design filters, biasing reduced-alphabet sequences toward chemically and biologically plausible class patterns before detailed physicochemical modeling or experimental optimization.

Motif constraints can be used when specific residue classes are required at selected positions. Let $M$ denote a set of motif requirements, where each element $(i,c_i^*) \in M$ indicates that position i should take residue class $c_i^*$. The corresponding penalty is

$$H_{motif} = \sum_{(i,c_i^*)\in M} (1 - x_{i,c_i^*})^2.$$

This term has the same mathematical form as the type constraint used for cyclization, but its interpretation is broader. It may encode a turn-promoting proline/glycine-like position, a cysteine-like residue required for later chemical elaboration, a charged residue at a solvent-facing position, or a hydrophobic class at a position expected to contribute to a binding interface. Motif terms therefore allow prior sequence knowledge or user-defined design rules to be added without changing the structure of the QUBO model.

Composition constraints control the overall number of residues belonging to selected residue-class groups. Let G denote a collection of class groups, where each group $g \subseteq C$ may represent hydrophobic, charged, polar, cysteine-like, or linker-compatible classes. If the desired number of positions assigned to group g is $n_g$, a quadratic composition penalty can be written as

$$H_{comp} = \sum_{g \in G} \left( \sum_{i=1}^{N} \sum_{c \in g} x_{i,c} - n_g \right)^2.$$

This term can enforce or bias global design features such as the number of charged positions, the number of hydrophobic positions, or the number of cysteine-like residues. When strict equality is not desired, the same idea can be relaxed by using lower and upper bounds or by assigning softer penalties around a preferred range.

Developability-related terms can be represented in a similar coarse-grained manner. For example, residue-class assignments may be penalized if they lead to excessive hydrophobicity, excessive net charge, poor solubility proxies, or chemically undesirable patterns. A generic developability contribution can be written as

$$H_{dev} = \alpha_{hyd} H_{hyd} + \alpha_{chg} H_{chg} + \alpha_{sol} H_{sol} + \alpha_{liab} H_{liab}.$$

Here $H_{hyd}$, $H_{chg}$, $H_{sol}$, and $H_{liab}$ denote coarse penalties associated with hydrophobicity, charge balance, solubility proxies, and sequence liabilities, respectively. These terms may be defined from simple class counts, position-specific penalties, or pairwise class patterns. For example, if $h_c$ denotes a hydrophobicity score assigned to residue class $c$, a simple hydrophobicity penalty can be written as

$$H_{hyd} = \left( \sum_{i=1}^{N} \sum_{c \in C} h_c x_{i,c} - h^* \right)^2,$$

where $h^*$ is a preferred class-level hydrophobicity target. Similar count- or score-based terms can be constructed for net charge or solubility proxies. These quantities should be interpreted as coarse design preferences rather than quantitative predictors of pharmacokinetics, permeability, or oral bioavailability. Only linear or quadratic proxy terms are included directly in the QUBO objective; more complex or higher-order property rules would require quadratization, approximation, or downstream filtering.

The motif, composition, and developability terms can therefore be summarized as

$$H_{rule} = \lambda_{motif} H_{motif} + \lambda_{comp} H_{comp} + \lambda_{dev} H_{dev}.$$

Together with the cyclization and target-binding terms, these additional rules allow the QUBO objective to balance structural feasibility, target awareness, and coarse drug-like design preferences within a single reduced-alphabet optimization framework.

## 3.5 Total Hamiltonian and Ising mapping

The preceding terms can now be assembled into a single reduced-alphabet QUBO objective. The one-hot term enforces sequence validity, the cyclization term encodes ring-formation requirements, the target term introduces target-binding compatibility, and the motif, composition, and developability terms collect additional sequence-level design preferences. The total Hamiltonian is written as

$$H_{total} = \lambda_{oh} H_{onehot} + \lambda_{cyc} H_{cyclization} + \lambda_{target} H_{target} + \lambda_{motif} H_{motif} + \lambda_{comp} H_{comp} + \lambda_{dev} H_{dev}.$$

Here, all $\lambda$ coefficients are non-negative weights that control the relative strengths of the corresponding Hamiltonian terms. In particular, $\lambda_{oh}$, $\lambda_{cyc}$, and $\lambda_{target}$ weight sequence validity, cyclization feasibility, and target-binding compatibility, whereas $\lambda_{motif}$, $\lambda_{comp}$, and $\lambda_{dev}$ weight motif, composition, and developability-related preferences. Equivalently, the last three terms can be grouped into $H_{rule}$, which collects motif, composition, and developability-related preferences. In practice, the one-hot and hard cyclization terms are usually assigned sufficiently large weights so that invalid bitstrings are energetically disfavored relative to valid candidates. Softer design preferences, such as hydrophobicity balance or target-interface preference, can be assigned lower weights when they are intended to bias rather than strictly constrain the search.

Because all variables are binary and all terms are at most quadratic in the variables $x_{i,c}$, the total objective can be written in standard QUBO form. To do so, the two-index variable $x_{i,c}$ is flattened into a single binary variable $q_\ell$, where $\ell$ indexes the pair $(i,c)$. There are NK binary variables in the one-hot encoding,

$$q_\ell \equiv x_{i,c}, \qquad \ell = 1,\ldots,NK.$$

The total Hamiltonian can then be written as

$$H_{total}(q) = q^T Q q + b^T q + C,$$

where $Q$ is the quadratic coefficient matrix, $b$ is the vector of linear coefficients, and $C$ is a constant energy offset. The entries of $Q$ and $b$ are determined by expanding the one-hot, cyclization, target-binding, motif, composition, and developability terms. The constant C does not affect the minimizer but may be retained for energy bookkeeping.

The QUBO representation can be converted into an Ising form by introducing spin variables

$$s_\ell \in \{-1, +1\}, \qquad q_\ell = \frac{1 + s_\ell}{2}.$$

Substituting this relation into the QUBO objective gives an Ising Hamiltonian of the form

$$H_{Ising}(s) = \sum_{\ell} h_{\ell}\, s_{\ell} + \sum_{\ell<m} J_{\ell m}\, s_{\ell} s_m + C',$$

where $h_{\ell}$ are local fields, $J_{\ell m}$ are pairwise couplings, and $C'$ is a shifted constant. This mapping does not change the underlying optimization problem; it only rewrites the binary objective in the spin-variable language used by Ising models and quantum annealing formulations.

The resulting formulation is solver-agnostic. The same $H_{total}$ can be minimized by classical heuristics, simulated annealing, quantum annealing, QAOA-like samplers, or other binary optimization procedures. A sampler returns low-energy bitstrings, which can then be checked for one-hot validity and decoded into residue-class sequences. Since the present work is a theoretical formulation, the Hamiltonian is the main object of interest; hardware implementation, parameter tuning, and empirical benchmarking are treated as downstream or future tasks.

This construction also clarifies the role of each modeling component. The one-hot term defines the search space, the cyclization term distinguishes cyclic peptide candidates from unconstrained linear sequences, the target term introduces binding-object awareness, and the rule term adds motif, composition, and coarse developability preferences. Candidate sequences with lower total energy are therefore interpreted as residue-class sequences that better satisfy the combined design requirements within the chosen reduced alphabet.

# 4. Decoding and Downstream Design Workflow

The QUBO/Ising formulation defined above produces low-energy binary configurations rather than final peptide molecules. A solver, whether classical or quantum-compatible, returns candidate bitstrings over the residue-class variables $x_{i,c}$. These bitstrings should be interpreted as class-level design proposals that satisfy, or approximately satisfy, the combined sequence-validity, cyclization, target-compatibility, and rule-based terms in the Hamiltonian. This readout mode is naturally aligned with sampler-like workflows, in which the computational output is a collection of measured or sampled bitstrings rather than a single deterministic molecular structure [16]. The role of the downstream workflow is therefore to convert these bitstrings into residue-class sequences, expand them into concrete amino-acid candidates, and evaluate them using more detailed structural or experimental procedures.

The first step is validity checking. Because the one-hot term enforces that each peptide position should be assigned to exactly one residue class, each returned bitstring is checked position by position. A bitstring is directly valid if, for every position $i$, exactly one variable $x_{i,c}$ equals 1. If no class or multiple classes are selected at a position, the candidate can either be discarded or repaired by selecting the class with the most favorable local contribution to the Hamiltonian. In strict implementations, invalid one-hot assignments are removed before downstream processing. In softer implementations, such candidates may be retained only if they can be repaired without violating hard cyclization constraints.

After sequence-validity checking, the candidate is evaluated against cyclization-related constraints. Type-like requirements are inspected first. For example, a disulfide design may require cysteine-like classes at two specified positions, whereas a stapled peptide may require linker-compatible residue classes at positions separated by a prescribed interval. Pairwise or topology-level cyclization terms are then checked to determine whether the selected class pattern is compatible with the intended ring-formation strategy. This step does not confirm atomistic ring closure; rather, it filters class-level sequences that are inconsistent with the selected cyclization mode. This distinction is important because cyclic peptide generation requires explicit topology and compatibility constraints that are not present in unconstrained linear peptide generation [10].

Once a candidate bitstring passes validity and cyclization checks, it is decoded into a residue-class sequence

$$z = (z_1, \ldots, z_N), \qquad z_i \in C,$$

where $z_i$ is the unique residue class selected at position i. This class sequence is the primary output of the present model. It defines a constrained region of sequence space rather than a single final peptide. If multiple low-energy bitstrings decode to the same residue-class sequence, they can be merged. If several distinct residue-class sequences have similar energies, they can be retained as a diverse candidate set for downstream expansion. Candidate ranking at this stage may use the total Hamiltonian energy as well as individual component scores, such as cyclization compatibility, target compatibility, and developability-related penalties.

The next step is amino-acid-level decoding. Each residue class is expanded into one or more concrete amino acids according to the chosen reduced alphabet. For example, a hydrophobic class may be expanded into residues such as Leu, Ile, Val, Phe, or related non-natural alternatives; a cysteine-like class may be expanded into cysteine or chemically equivalent thiol-containing building blocks; and a linker-compatible class may be mapped to residues suitable for a selected stapling or macrocyclization chemistry. This step can be deterministic when a class corresponds to a required residue, as in a fixed disulfide constraint, or stochastic when several amino acids are chemically plausible. Target-conditioned sequence-generation tools may also be used at this stage to refine residue-level choices within the class-level scaffold [11].

The amino-acid expansion step should remain aware of the selected cyclization mode. For head-to-tail cyclization, the decoded terminal residues must be compatible with downstream amide-bond formation or chemical activation. For disulfide cyclization, the relevant cysteine-like positions must be converted into residues that can form the intended bridge. For stapled or bicyclic designs, residue identities, spacing, and linker chemistry must be jointly checked. In this sense, the QUBO output serves as a class-level scaffold that reduces the search space, while detailed chemical construction is deferred to a later design layer.

After concrete sequences are generated, structural validation can be applied. Candidate peptides may be subjected to conformer generation, cyclic-closure modeling, docking, molecular-dynamics simulation, or other structure-based filters. Prior cyclic peptide QUBO work has considered docking as an optimization problem for a given cyclic peptide [14], whereas the present workflow places docking downstream of class-sequence generation. Similarly, quantum-assisted peptide design has used reduced encodings as part of a broader multiscale candidate-generation framework [13]. The present model follows the same general principle that a binary optimization layer should be connected to downstream molecular evaluation rather than treated as a complete replacement for structural modeling.

The final stage is candidate prioritization. Low-energy residue-class sequences are not automatically the best experimental candidates; they are candidates that satisfy the encoded assumptions of the Hamiltonian. Therefore, downstream ranking should combine multiple criteria, including total QUBO energy, satisfaction of hard cyclization rules, target-compatibility scores, sequence diversity, amino-acid-level feasibility, predicted structural stability, docking or interface scores, and developability-related filters. Candidates that remain favorable across these layers can then be selected for more detailed simulation or experimental testing. Thus, the proposed QUBO/Ising formulation should be viewed as an early-stage search-space reduction and prioritization tool. It provides a principled route from combinatorial class-level optimization to concrete cyclic peptide design, while leaving atomistic modeling and empirical validation to subsequent stages.

# 5. Model Properties and Relationship to Existing Frameworks

## 5.1 Modularity of the Hamiltonian

A central property of the proposed formulation is that the design objective is modular. The total Hamiltonian is not constructed as a single monolithic scoring function, but as a weighted sum of interpretable terms. The one-hot term enforces residue-class assignment validity; the cyclization term encodes ring-formation requirements; the target term introduces class-level compatibility with a fixed binding environment; and the motif, composition, and developability terms represent additional user-defined design rules. This organization allows different biological or chemical requirements to be added, removed, or reweighted without changing the basic binary optimization structure.

This modularity is important because cyclic peptide design is not governed by a single constraint. A candidate may be chemically compatible with a ring-closure strategy but poorly matched to the target interface, or it may contain a favorable target-facing pattern while violating composition or developability preferences. Expressing these requirements as separate Hamiltonian terms makes the trade-offs explicit. It also allows the framework to support different design scenarios. For example, a disulfide-constrained peptide, a head-to-tail macrocycle, and a stapled peptide may use the same class-assignment variables but different cyclization penalties. Similarly, target-related terms can be omitted for target-agnostic library generation or strengthened when binding-site information is available.

From an optimization perspective, this structure follows the standard practice of converting constrained binary problems into unconstrained objectives by adding penalty terms. Such penalty-based reformulations are common in pseudo-Boolean and UBQP/QUBO modeling, where hard constraints and soft preferences are represented within a single binary objective [17–19]. The present model applies this general strategy to reduced-alphabet cyclic peptide design.

## 5.2 Solver-agnostic formulation

The QUBO/Ising representation should be interpreted as a modeling language rather than a commitment to a specific solver. Once the cyclic peptide design problem is expressed as a binary objective, it can in principle be explored by different optimization procedures. These may include deterministic or stochastic classical heuristics, simulated annealing, tabu-style search, quantum annealing, or gate-model sampling workflows. The formulation therefore separates the peptide-design model from the computational device used to search it.

This distinction is important for avoiding an overstatement of the role of quantum computing. The model is quantum-compatible in the sense that it can be written in QUBO or Ising form, but this does not imply quantum advantage. Classical simulated annealing is itself grounded in the analogy between statistical mechanics and combinatorial optimization [20]. Quantum annealing extends this type of search by using quantum fluctuations, often represented through a transverse-field Ising model, as a mechanism for transitions between classical states [21]. Gate-model approaches may also sample bitstrings from parameterized quantum circuits, but their performance depends on circuit depth, noise, ansatz choice, and classical optimization strategy.

For this reason, the present work does not claim that any specific solver will outperform classical alternatives for cyclic peptide design. Instead, it defines a solver-agnostic binary formulation. This is consistent with prior QUBO and Ising literature, where the same binary objective can be treated as a mathematical optimization problem, a classical heuristic search problem, or a quantum-compatible Hamiltonian [17–19]. In practical use, solver choice should be treated as an implementation question and evaluated empirically.

## 5.3 Scaling and reduced-alphabet representation

The reduced-alphabet assumption is both a strength and a limitation of the framework. Full amino-acid identity is chemically richer than residue-class identity: in atomistic docking, each side chain has distinct geometry, charge distribution, hydrogen-bonding capacity, rotameric behavior, and solvation properties. Therefore, the reduced alphabet used here should not be interpreted as a replacement for full 20-amino-acid modeling. It is instead a working representation for early-stage class-level design-space reduction.

Reduced amino-acid alphabets have precedent in protein modeling and fold-recognition studies. Previous work has shown that simplified alphabets can preserve useful structural information, although excessive alphabet reduction leads to information loss [22]. Other studies have also reported that reduced alphabets can improve sensitivity and selectivity in fold-assignment settings, especially when structural similarity is present despite weak sequence identity [23]. These results do not prove that a reduced alphabet is sufficient for cyclic peptide docking, but they support the broader idea that amino-acid grouping can be a meaningful coarse-grained representation.

In the present formulation, the reduced alphabet also affects binary encoding cost. In the logical one-hot QUBO formulation, the number of peptide variables scales as $NK$, where $N$ is the peptide length and $K$ is the number of residue classes. In a compact implementation, the code length scales as $\lceil \log_2 K \rceil$, although arbitrary class-score lookup tables may require quadratization or approximation. Thus, the choice of $K$ involves a trade-off between chemical resolution, interaction-pattern preservation, and encoding cost. The MJ-based analysis supporting the default eight-class working alphabet is summarized in Section 5.4.

## 5.4 MJ-based reduced-alphabet results

To support the default reduced alphabet, we performed an MJ-based clustering analysis of the 20 natural amino acids. The purpose of this analysis was not to define a new biochemical taxonomy, but to test whether a smaller set of residue classes could preserve the coarse interaction structure contained in the original Miyazawa–Jernigan (MJ) residue–residue interaction matrix [15]. Each amino acid was represented by its 20-dimensional MJ interaction profile, corresponding to one row of the MJ matrix. Reduced alphabets with $K = 2$ to $K = 12$ classes were evaluated. For each $K$, amino acids were clustered by K-means using the original MJ profiles without feature scaling, and silhouette scores were computed using Euclidean distances.

For each grouping, a grouped $K \times K$ MJ matrix was constructed by averaging the original MJ entries between amino acids assigned to the corresponding residue groups. This grouped matrix was then expanded back to a reconstructed $20 \times 20$ matrix. Reconstruction quality was evaluated by root-mean-square error (RMSE), Pearson correlation, Spearman correlation, and retention of the lowest-energy 10% amino-acid pairs. These metrics were used to assess whether the reduced representation preserved coarse interaction patterns rather than full atomistic specificity. The resulting clustering structure and fidelity–compact encoding trade-off are summarized in Figure 2.

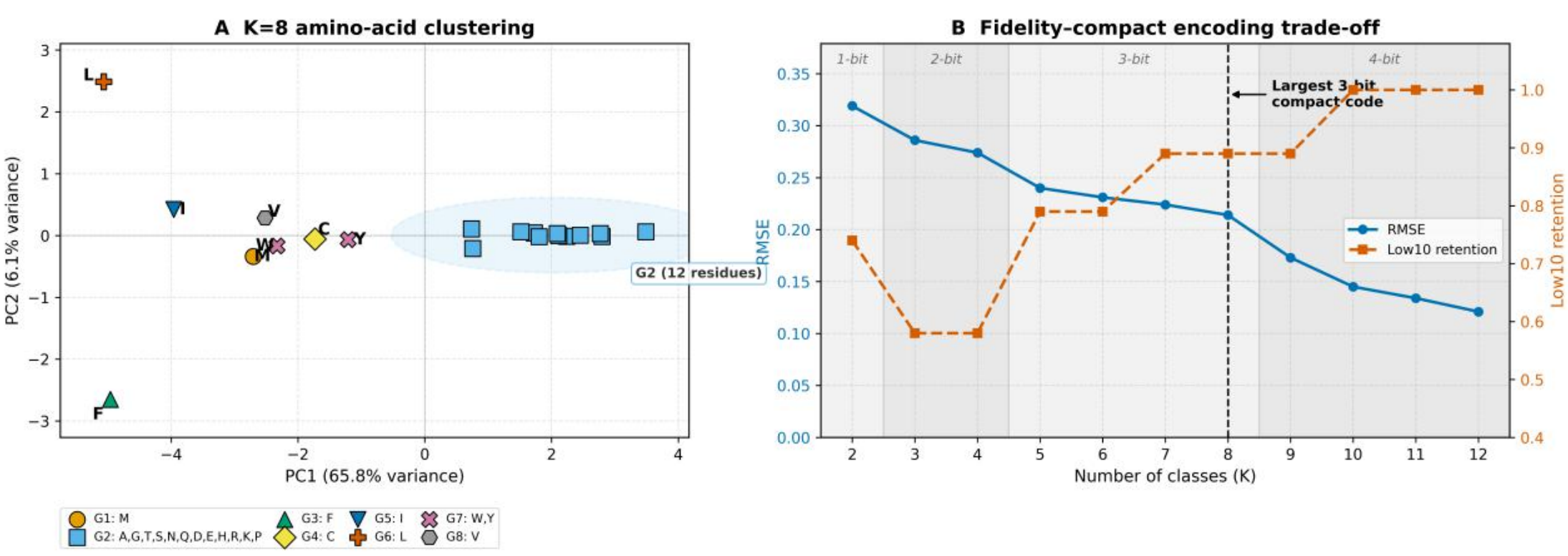

**Figure 2. MJ-based clustering and reduced-alphabet selection for compact binary encoding.** (A) Principal component analysis of the 20 amino-acid interaction profiles derived from the Miyazawa – Jernigan (MJ) residue – residue interaction matrix, colored according to the $K = 8$ clustering result. Each point represents one amino acid, and proximity reflects similarity in MJ contact-energy profiles rather than atomistic or structural distance. The large G2 group contains 12 residues and reflects a broad interaction-profile cluster rather than a conventional physicochemical class. (B) Fidelity – compact encoding trade-off across reduced alphabet sizes. RMSE measures reconstruction error between the original MJ matrix and the grouped reconstruction, whereas Low10 retention measures preservation of the lowest-energy 10% amino-acid pairs. Gray bands indicate compact-code bit regimes, and the dashed vertical line marks $K = 8$ as the largest reduced alphabet that remains within a 3-bit compact code.

Figure 2A visualizes the $K = 8$ MJ interaction-profile grouping, while Figure 2B shows the corresponding fidelity–compact encoding trade-off. The PCA map should be interpreted only as a two-dimensional summary of MJ-profile similarity. Together, the two panels support treating $K = 8$ as a compact working alphabet rather than as a conventional biochemical classification. The full $K = 2$ to $K = 12$ clustering and reconstruction metrics are reported in Table 2.

**Table 2. Clustering and fidelity – encoding metrics for MJ-matrix clustering across reduced amino-acid alphabet sizes.**

| Number of classes K | Compact-code bits per position | Silhouette | RMSE | Pearson | Spearman | Low-energy pair retention |
|---|---|---|---|---|---|---|
| 2 | 1 | 0.490 | 0.319 | 0.910 | 0.903 | 0.74 |
| 3 | 2 | 0.419 | 0.286 | 0.932 | 0.932 | 0.58 |
| 4 | 2 | 0.417 | 0.274 | 0.933 | 0.933 | 0.58 |

| Number of classes K | Compact-code bits per position | Silhouette | RMSE | Pearson | Spearman | Low-energy pair retention |
|---|---|---|---|---|---|---|
| 5 | 3 | 0.409 | 0.240 | 0.945 | 0.936 | 0.79 |
| 6 | 3 | 0.397 | 0.231 | 0.948 | 0.934 | 0.79 |
| 7 | 3 | 0.385 | 0.224 | 0.951 | 0.931 | 0.89 |
| 8 | 3 | 0.360 | 0.214 | 0.955 | 0.931 | 0.89 |
| 9 | 4 | 0.131 | 0.173 | 0.971 | 0.964 | 0.89 |
| 10 | 4 | 0.127 | 0.145 | 0.980 | 0.967 | 1.00 |
| 11 | 4 | 0.114 | 0.134 | 0.982 | 0.965 | 1.00 |
| 12 | 4 | 0.089 | 0.121 | 0.986 | 0.970 | 1.00 |

Note. Compact-code bits per position are computed as $\lceil \log_2 K \rceil$ and are reported only as a compact-encoding reference; the main QUBO formulation in this work uses one-hot class indicators. Silhouette scores were computed using Euclidean distances on the original MJ interaction profiles. RMSE, Pearson correlation, and Spearman correlation measure reconstruction fidelity between the original MJ matrix and the grouped reconstruction. Low-energy pair retention measures the overlap between the lowest-energy 10% amino-acid pairs in the original MJ matrix and those recovered after class-level reconstruction.

The quantitative results further support this trade-off. As shown in Table 2, increasing $K$ reduces reconstruction error and improves overall matrix fidelity, but the compact-code cost also changes across different $K$ regimes. $K = 8$ was selected because it is the largest reduced alphabet that remains within a 3-bit compact code while retaining substantial MJ interaction structure and preserving most of the strongest favorable contact pairs. Larger alphabets, such as $K = 9$ and $K = 10$, improve RMSE and correlation values, but they enter the 4-bit compact-code regime.

Thus, the eight-class alphabet is used here as a resource-aware working representation for early QUBO-level filtering, rather than as a biologically optimal amino-acid classification. MJ-type contact energies are treated as coarse interaction priors, not as all-atom energy functions or docking scores. Downstream amino-acid decoding, structural modeling, docking, molecular dynamics, and experimental validation remain necessary.

## 5.5 Relationship to CP-Composer

The proposed framework is related to CP-Composer in its use of composable constraints, but the modeling goal is different. CP-Composer is a generative geometric framework for cyclic peptide generation. It decomposes complex cyclization patterns into type and distance constraints and uses these constraints to guide cyclic peptide generation through a neural model [10]. Its main contribution lies in zero-shot cyclic peptide generation using composable geometric conditioning.

The present work borrows the idea that cyclization can be decomposed into elementary constraints, but it does not use a diffusion model, does not generate atomistic peptide conformations, and does not learn a neural generative distribution. Instead, it translates cyclization-relevant requirements into QUBO-compatible residue-class penalties. In this sense, CP-Composer motivates the constraint vocabulary, whereas the present work reformulates related design requirements as a reduced-alphabet binary optimization problem.

## 5.6 Relationship to quantum peptide binder design

The closest reference point on the quantum-design side is the quantum-assisted de novo peptide binder framework of Meuser et al. [13]. That work uses reduced molecular encodings and quantum-compatible optimization to design peptide binders, including sequence and binding-pose considerations. It is therefore highly relevant as a demonstration that peptide design can be connected to binary optimization and quantum annealing workflows.

However, the present formulation differs in task definition. It focuses on cyclic peptide class-sequence generation under explicit cyclization and target-compatibility constraints. The optimized variables are residue-class assignments on the peptide, while the target is treated as a fixed external environment. The model does not attempt to jointly optimize atomistic peptide sequence and binding pose. Its purpose is to generate constrained class-level candidates that can later be decoded and validated by downstream structural methods.

## 5.7 Relationship to cyclic peptide QUBO docking

Brubaker et al. provide an important cyclic peptide QUBO reference because they formulate lattice-based cyclic peptide docking with both peptide cyclization and peptide–protein docking objectives [14]. That work demonstrates that cyclic topology and target interaction can be placed within a QUBO-compatible framework. It also highlights practical scaling challenges, since their QUBO approach could solve only relatively small cyclic peptide docking instances before becoming difficult to scale.

The present work uses a different level of abstraction. Brubaker et al. address fixed-sequence cyclic peptide docking on a lattice, whereas this paper addresses residue-class sequence generation before atomistic docking. The goal is not to find a docking pose for a given cyclic peptide, but to prioritize class-level cyclic peptide patterns that satisfy cyclization and target-related design rules. Thus, the two approaches are complementary: docking-oriented

QUBO models operate closer to structure prediction, while the present model operates earlier in the design pipeline.

### 5.8 Summary of positioning

Overall, the proposed model is best understood as a reduced-alphabet, constraint-driven QUBO/Ising formulation for early-stage cyclic peptide design. Its contribution is not a new peptide force field, a new docking algorithm, or a claim of quantum advantage. Rather, it provides a modular binary optimization framework that integrates sequence validity, cyclization feasibility, target compatibility, motif constraints, composition control, and coarse developability preferences. By combining this structure with an MJ-supported eight-class working alphabet, the model offers a resource-aware route for generating cyclic peptide class-sequence candidates that can be expanded and validated downstream.

## 6. Discussion

The present work proposes a reduced-alphabet QUBO/Ising formulation for target-aware cyclic peptide class-sequence design. Its main contribution is not the introduction of a new peptide force field, a new docking algorithm, or a new quantum solver, but the organization of cyclic peptide design requirements into a modular binary optimization framework. In this framework, sequence validity, cyclization feasibility, target compatibility, motif requirements, composition control, and coarse developability preferences are expressed as Hamiltonian terms that can be combined within a single objective function. This provides a formal route for translating cyclic peptide design rules into an optimization problem that is compatible with both classical and quantum-inspired binary solvers [17–19].

The formulation is intended for early-stage design-space reduction. In conventional peptide design, the combinatorial space grows rapidly with peptide length, especially when all 20 natural amino acids and possible non-natural building blocks are considered. Cyclic peptide design further narrows and structures this space because candidate sequences must satisfy ring-formation rules, residue-type requirements, spacing patterns, and, in target-aware settings, interface compatibility constraints [7,10]. The proposed QUBO/Ising formulation does not attempt to search the full atomistic sequence–structure–binding space directly. Instead, it first identifies residue-class patterns that satisfy encoded design assumptions. These class-level candidates can then be expanded into concrete amino-acid sequences and evaluated by downstream structural and experimental methods.

A key feature of the framework is the use of a reduced amino-acid alphabet. This choice is useful from an optimization perspective because it reduces the class space and, in compact implementations, lowers the number of bits required to represent each peptide position. However, it is also one of the main approximations of the model. Full amino-acid identity contains side-chain geometry, rotameric behavior, hydrogen-bonding patterns, charge distribution, solvation effects, and chemical reactivity that cannot be fully captured by residue classes. The eight-class alphabet used here should therefore not be interpreted as a biologically optimal amino-acid taxonomy or as a replacement for atomistic modeling. It is a resource-aware working alphabet motivated by MJ interaction-profile clustering and by the need to balance energy fidelity with binary encoding cost.

The MJ clustering analysis provides partial support for this reduced representation. In the tested grouping scheme, $K = 8$ preserved substantial MJ interaction structure while remaining within a three-bit compact code. Larger alphabets such as $K = 9$ or $K = 10$ showed better RMSE and correlation values, but required a transition to four compact bits per position. This makes $K = 8$ a practical compromise rather than a universal optimum. Prior studies of reduced amino-acid alphabets also support the broader idea that amino-acid grouping can preserve useful structural or sequence information, while also warning that excessive compression causes information loss [22,23]. In the present model, this means that the reduced alphabet is appropriate for coarse class-level filtering, but final peptide selection must recover amino-acid specificity during downstream decoding.

The scoring terms used in the Hamiltonian should also be interpreted with care. MJ-type residue–residue potentials provide a convenient source of coarse contact preferences, but they are not all-atom force fields and do not represent full cyclic peptide energetics [15]. Similarly, the target compatibility term is not a docking score or a binding free-energy estimator. It uses fixed target priors, such as residue classes, contact weights, or interface annotations, to bias peptide class assignments toward a hypothesized binding environment. It does not model conformational entropy, induced fit, water-mediated interactions, side-chain packing, ring strain, or solvation. The same caution applies to the developability terms: class-level hydrophobicity, charge, solubility, or liability penalties can guide early filtering, but they should not be treated as quantitative predictors of pharmacokinetics, permeability, toxicity, or oral bioavailability.

Another limitation is the absence of empirical validation in the current work. The formulation specifies how cyclic peptide design rules can be encoded as a QUBO/Ising objective, but it does not yet test whether the resulting low-energy class sequences produce experimentally useful cyclic peptides. This is consistent with the scope of a theoretical formulation paper, but it also defines the next level of work. Future validation should test whether low-energy bitstrings correlate with downstream peptide quality after amino-acid decoding, conformer generation, cyclic closure modeling, docking, molecular dynamics, and experimental screening. Without such validation, QUBO energy should be interpreted only as a score under the assumed model, not as a direct proxy for biological activity.

The solver-agnostic nature of the formulation is both useful and deliberately conservative. Because the objective can be written in QUBO or Ising form, it may be explored using classical heuristics, simulated annealing, quantum annealing, QAOA-like samplers, or other binary optimization procedures [17–21]. However, compatibility with quantum or quantum-inspired solvers does not imply quantum advantage. Existing peptide and cyclic peptide QUBO studies already show that scaling, embedding, constraint handling, and solver performance remain important practical issues [12–14]. The present formulation should therefore be evaluated by comparing different solvers on the same Hamiltonian, with attention to solution quality, feasibility rate, diversity, runtime, and scalability.

Future work should proceed in several stages. First, the eight-class alphabet should be tested beyond MJ matrix reconstruction by comparing K = 8, K = 9, and K = 10 alphabets on actual cyclic peptide design tasks. Relevant criteria include class-sequence diversity, constraint satisfaction, amino-acid decoding success, docking scores, conformational stability, and target-interface compatibility. Second, class-to-amino-acid decoding strategies should be developed so that residue-class scaffolds can be converted into chemically specific peptide candidates while preserving cyclization chemistry. Third, candidate sequences should be evaluated through conformer generation, cyclization-aware structural reconstruction, docking, molecular-dynamics simulation, and binding-interface analysis. Fourth, penalty weights and solver settings should be systematically benchmarked, since the balance between hard constraints and soft preferences will affect the resulting candidate set. Finally, selected candidates should be synthesized and tested experimentally for binding, stability, protease resistance, and other developability-related properties. Such feedback would allow the Hamiltonian terms and reduced-alphabet definitions to be recalibrated from empirical data.

Overall, the value of the proposed framework lies in its explicit separation of modeling layers. The QUBO/Ising objective defines a constrained class-level search problem; decoding converts class patterns into amino-acid candidates; structural modeling evaluates physical plausibility; and experimental testing determines biological relevance. This layered interpretation avoids overstating the current formulation while preserving its utility as a bridge between cyclic peptide design rules and combinatorial optimization.

# 7. Conclusion

This work presents a reduced-alphabet QUBO/Ising formulation for target-aware cyclic peptide class-sequence design. The central aim is to express cyclic peptide design as a constrained binary optimization problem, rather than as direct atomistic structure prediction, docking, or binding free-energy estimation. By representing each peptide position through residue-class assignments, the formulation provides a compact design layer in which cyclic peptide sequence validity, ring-formation requirements, target-related compatibility, and additional sequence-level rules can be handled within a unified Hamiltonian framework.

The proposed model is modular by construction. One-hot penalties define valid residue-class sequences, cyclization terms distinguish cyclic peptide candidates from unconstrained linear sequences, target-compatibility terms introduce information from a fixed binding environment, and motif, composition, and developability terms allow additional design preferences to be incorporated. The eight-class reduced alphabet used as the default representation is motivated by MJ interaction-profile clustering and should be viewed as a resource-aware working alphabet rather than a biologically complete amino-acid taxonomy. The QUBO/Ising form further makes the model solver-agnostic: the same objective can be explored by classical heuristics, simulated annealing, quantum annealing, QAOA-like samplers, or other binary optimization procedures, without implying quantum advantage.

The output of the framework is not a final therapeutic candidate, but a set of low-energy residue-class sequences that satisfy the assumptions encoded in the Hamiltonian. These candidates require amino-acid-level decoding, cyclization-aware chemical construction, structural modeling, docking, molecular-dynamics simulation, and experimental validation before biological relevance can be assessed. The value of the formulation therefore lies in its role as an early-stage search-space reduction and prioritization layer. It establishes a theoretical bridge between cyclic peptide design rules and combinatorial optimization, and provides a basis for future work on solver benchmarking, alphabet refinement, downstream structural validation, and data-calibrated cyclic peptide design.